\documentclass[conference]{IEEEtrancr}
\IEEEoverridecommandlockouts

\usepackage[T1]{fontenc}
\usepackage{cite}
\usepackage{amsmath,amssymb,amsfonts}
\usepackage{algorithmic}
\usepackage{graphicx}
\usepackage{textcomp}
\usepackage{xcolor}
\usepackage{hyperref}
\usepackage{color}

\usepackage{enumitem}
\usepackage{multicol}
\usepackage{cleveref}
\usepackage{booktabs}
\usepackage{tabularx}
\usepackage{url}
\usepackage{adjustbox}
\usepackage{xcolor}

\def\BibTeX{{\rm B\kern-.05em{\sc i\kern-.025em b}\kern-.08em
    T\kern-.1667em\lower.7ex\hbox{E}\kern-.125emX}}
\begin{document}

\title{Do (Not) Tell Me About My Insecurities: Assessing the Status Quo of Coordinated Vulnerability Disclosure in Germany Amid New EU Cybersecurity Regulations}

\newif\ifisanonymized
\isanonymizedfalse
\newif\ifnotisanonymized
\ifisanonymized
  \notisanonymizedfalse
\else
  \notisanonymizedtrue
\fi

\ifnotisanonymized
\author{
\IEEEauthorblockN{Sebastian Neef}
\IEEEauthorblockA{\textit{Technische Universität Berlin} \\
Berlin, Germany \\
0000-0003-3055-0823}
\and
\IEEEauthorblockN{Cenk Schlunke}
\IEEEauthorblockA{\textit{Technische Universität Berlin} \\
Berlin, Germany\\
0009-0002-3142-5776}
\and
\IEEEauthorblockN{Anne Hennig}
\IEEEauthorblockA{\textit{Karlsruhe Institute of Technology} \\
Karlsruhe, Germany \\
0000-0002-6964-589X}
}
\else
\author{
\IEEEauthorblockN{Anonymized}
\IEEEauthorblockA{\textit{Anonymized} \\
\textit{Anonymized} \\
Anonymized \\
Anonymized}
}
\fi 

\newcommand{\rev}[1]{#1}
\newcommand{\sq}[1]{\textit{#1}}

\newcommand{\rqone}{What is the status quo of CVD programs in German DAX companies and how does it change over time?}
\newcommand{\rqtwo}{What are the challenges and experiences in launching and running a CVD program for the German DAX companies?}
\newcommand{\rqthree}{What can be learned from surveying the German DAX companies and how can the insights benefit SMEs or policy makers?}
\newcommand{\primarysurvey}{\emph{Primary Survey}}
\newcommand{\secondarysurvey}{\emph{Secondary Survey}}

\maketitle

\begin{abstract}
In our increasingly interconnected world, good IT security practices are necessary to prevent vulnerabilities and data breaches. 
Providing security contacts, e.g., via Coordinated Vulnerability Disclosure (CVD) programs or security.txt files, is an important practice for businesses to facilitate vulnerability reporting by external parties.
As part of a longitudinal study, we analyzed the adoption of, as well as the challenges and experiences with, CVD programs among the 40 companies listed on Germany's DAX (the country's primary stock market index).
In addition to monitoring publicly available information about their CVD programs, we sent out questionnaires via email and postal mail in 2023 and 2025, and received answers from 20\% of the companies. 
The adoption rates show a significant increase from 50\% (2023) to over 90\% (2025), with ten new CVD programs and 25 new security.txt files now available.
The survey answers reveal that, for example, legal obligations (e.g., NIS2 and CRA) drive the adoption of CVD practices, but a lack of (human) resources and varying report quality are considered drawbacks.
As the first study to survey 40 German stock market index (DAX) companies on their CVD practices, our results can help foster the adoption and understanding of security programs among SMEs and other companies, and provide policymakers with insights into practical challenges and industry experiences. 
\end{abstract}


\thanks{© 2025 IEEE. Personal use of this material is permitted. Permission from
  IEEE must be obtained for all other uses, in any current or future media,
  including reprinting/republishing this material for advertising or
  promotional purposes, creating new collective works, for resale or
  redistribution to servers or lists, or reuse of any copyrighted component
  of this work in other works. Published in: 2025 European Symposium on Usable Security (EuroUSEC). DOI: \href{https://ieeexplore.ieee.org/abstract/document/11300384}{10.1109/EuroUSEC69254.2025.00020}}

\section{Introduction}
\label{sec:intro}

We live in an era of ever-increasing digitalization, interconnection, and complexity of systems and networks that power our modern world. 
With this, the risk of cybersecurity attacks and related threats also rises. Security breaches are reported regularly, some targeting high-value targets and resulting in large exposures of personal data (as reported by HaveIBeenPwned.com \cite{haveibeenpwnedHaveBeen}).
Thus, the identification and mitigation of (severe) vulnerabilities is necessary to prevent such breaches. 
While companies tend to have internal security teams and external security contractors perform security audits of their infrastructure and software at specific points in time, these audits can be costly and do not guarantee the discovery of all security issues.  

In order to receive valuable information about potential security issues at any time from anyone (e.g., security researchers or ethical hackers), so-called ``Coordinated Vulnerability Disclosure (CVD)'' programs have emerged. These programs define a framework of rules for what can be tested and how discovered security issues can be safely reported by external parties. 
CVD platforms (e.g., HackerOne \cite{hackeroneHackerOne}, Bugcrowd \cite{bugcrowdManagedBounty}) are frequently used by ethical hackers to report (severe) vulnerabilities. 
To report vulnerabilities other than via CVD platforms effectively, companies need to provide appropriate notification channels and contact information. Large-scale notification campaigns \cite{Durumeric.2014, Kuehrer.2014, Li.20167ep, Li.2016, Lone.2022, Maass.2021w1o, Maass.2021, Nosyk.2023, Rodriguez.2021, Stock.2018, Stock.2016, StopBadware.2012, Vasek.2012, Zeng.2019, Çetin.2017, Çetin.2018, Cetin.2019ndss, Çetin.2019, Çetin.2016} or attempts to deploy a coordinated process for vulnerability reports by researchers \cite{Chen.2024, Moura.2023} have shown that these notification channels either do not exist or that companies are nevertheless hesitant to react to the disclosures. 

However, upcoming legal changes in the European Union (EU) may require companies to provide reporting channels. \rev{One such change is the Cyber Resilience Act (CRA), which came into force on December 11, 2024, and must be adopted within 21 to 36 months. Another is the Network and Information Systems (NIS2) Directive, which came into force on January 16, 2021, and was supposed to be transposed into national law by October 17, 2024.} 
According to Article 21 of the Network and Information Systems (NIS2) Directive (No. 2022/2555~\cite{NIS2.2022}), establishing reporting channels may become an obligation for companies as an appropriate cybersecurity risk management measure.

Therefore, we asked ourselves how prepared companies that might fall under the NIS2 Directive are for coordinated vulnerability disclosure. 
\rev{
Closest to our work is the study by Walshe et al.~\cite{Walshe.2022}, which examined companies' experiences with CVD programs in 2022. They recruited companies for their study from several bug bounty platforms, meaning that all of them already had a CVD program in place.}

\rev{At the time of our first study, not all companies have adopted CVD practices (e.g., programs on dedicated platforms). Thus, we build upon their work by trying to understand why some companies have not (yet) implemented such programs, and how CVD adoption rates change in light of the new EU cybersecurity provisions over time between 2023 and 2025.
Furthermore, by gathering insights from companies with and without a CVD program, along with their respective challenges and experiences, we aim to derive recommendations and lessons learned that can benefit other businesses planning to adopt CVD practices and policymakers seeking to enhance vulnerability management.
For this reason, we focus on German companies, as Germany is one of the countries that has not yet incorporated the NIS2 Directive into national law at the time of writing, but where companies may expect it to happen in the near future. 
} 

\rev{Specifically, this study aims to extend the body of literature by answering the following research questions related to the adoption rates and processes of CVD programs among companies, as well as their motivations for or against implementing such programs.} 

\begin{enumerate}[leftmargin=1.1cm,labelwidth=.9cm,labelsep=.1cm,align=left,label=\textbf{RQ\arabic*:}]
    \item \rqone
    \item \rqtwo
\end{enumerate}

\rev{To answer our research questions, we analyzed publicly available information on the CVD programs of the 40 most highly valued publicly traded companies (according to the DAX stock market index) to assess their adoption in a first step. 
In a second step, we conducted two surveys to obtain further information on the introduction of and experiences with their CVD processes, in order to draw recommendations and lessons learned.}
Finally, we seek to raise awareness and encourage the adoption of CVD programs.

\section{Background and Related Work}
\label{sec:related-work}

\subsection{Coordinated Vulnerability Disclosure (CVD)}
Coordinated Vulnerability Disclosure (CVD) programs have become necessary to provide organized processes for the disclosure of -- sometimes severe -- vulnerabilities. 
As described by Ruohonen and Timmers \cite{Ruohonen.2024}, CVD programs arose from disclosure practices that were either direct disclosure (the party discovering a vulnerability directly notifies the party being affected without involving further third parties, such as public authorities) or full disclosure (the party discovering the vulnerability makes it public without notifying the party being affected). 

CVD programs 
allow the party affected by the vulnerability to be notified first and given time to publish security patches before the vulnerability is made public in a coordinated disclosure process. Thus, as the name suggests, both parties cooperate in a coordinated process to handle the vulnerability report and its disclosure.
However, as described in~\Cref{sec:intro}, for disclosers who report vulnerabilities, companies need to provide proper communication channels. Related work~\cite{Durumeric.2014, Kuehrer.2014, Li.20167ep, Li.2016, Lone.2022, Maass.2021w1o, Maass.2021, Nosyk.2023, Rodriguez.2021, Stock.2018, Stock.2016, StopBadware.2012, Vasek.2012, Zeng.2019, Çetin.2017, Çetin.2018, Cetin.2019ndss, Çetin.2019, Çetin.2016, Chen.2024, Moura.2023} has shown that this is not always the case. We will discuss this matter in more detail in~\Cref{sec:related-work:handling}.

One easy solution to facilitate vulnerability reporting is provided by RFC 9116 \cite{rfc9116}, which was standardized in 2022. 
It proposes a human-readable and machine-parsable text file (``security.txt'') that web administrators can place at \texttt{/.well-known/security.txt} or \texttt{/security.txt} (legacy) paths on their website. The file must be served over an SSL/TLS-secured connection (HTTPS) with a ``text/plain'' content type to preserve the authenticity of the data.
The file contains useful information for vulnerability reporters as structured key-value pairs, which help establish contact to report a security vulnerability. 
For example, required fields are \emph{Contact:} for details such as email addresses for a security team and an expiration date (\emph{Expires:}). Additional information placed in that file can include \emph{Encryption:} for information about secure communication (e.g., encrypted with PGP), a link to a vulnerability disclosure policy in \emph{Policy:} to communicate a CVD program's scope and rules, job opportunities (\emph{Hiring:}), a link to acknowledgments (\emph{Acknowledgments:}), or others.
However, the standard still lacks widespread adoption, as shown by large-scale analyses
\cite{Poteat.2021,hilbig.2023,Findlay.2022}.

\subsection{Legal Changes Affecting CVD}
Upcoming legal changes in the European Union (EU) may require companies to provide reporting channels.
The ISO/IEC 29147:2018~\cite{Iso.2018} of the International Organization for Standardization (ISO) provides guidelines for the process of vulnerability disclosure. Specifically, it offers guidelines for organizations on how to receive, review, and manage vulnerability reports effectively. These guidelines have been adapted and refined, for example, by governments to provide coordinated processes for their countries~\cite{ENISA.2022,BSI.2022,CMU.2017}. With the NIS2 Directive, the Cyber Resilience Act (CRA), and the General Data Protection Regulation (GDPR)\footnote{According to Articles 33 and 34 GDPR, supervisory authorities must be informed if a vulnerability leads to the loss or compromise of personal data.}, the European Union has made vulnerability disclosure, the reporting of data breaches, and the establishment of CVD processes a legal requirement~\cite{Ruohonen.2024,Vandezande.2024,Vostoupal.2024,Schmitz.2021}.

Despite the growing importance of vulnerability management within the new EU legal changes (especially the \rev{NIS2 rules}) and the adoption of CVD practices in many other countries~\cite{Pil.2023}, the current status quo of CVD programs among German companies remains largely unexplored. Thus, we derived our first research question: \textit{\textbf{RQ1:} \rqone}

\subsection{Handling of CVD}
\label{sec:related-work:handling}
Providing information on CVD programs may be laborious for companies at first, but missing specifications for CVD processes can cause further problems, as Kranenbarg et al. \cite{Kranenbarg.2018} describe: disclosers might not know whom to contact, and vulnerability disclosure receivers might not have the appropriate processes or rules in place to handle the reports. This can lead to a frustrating experience on both sides~\cite{Kranenbarg.2018}.

From a discloser's perspective, vulnerability disclosure can be time-consuming and demanding \cite{Moura.2023}, and those who disclose vulnerabilities might face negative reactions \cite{Stoever.2023,Maass.2021,Moura.2023}. 
Furthermore, several studies have reported problems reaching out to the appropriate stakeholders (\cite{Maass.2021,Stock.2018,Vasek.2012,Utz.2023,Chen.2024}). While Poteat et al. showed that the adoption of security.txt files \cite{rfc9116}, which are supposed to provide contact information for those who want to report vulnerabilities, is limited \cite{Poteat.2021}, even when a proper security contact exists, recipients might be difficult to reach \cite{Hove.2023}.
To offset the problem of non-reachability, CVD platforms (i.e., Bugcrowd \cite{bugcrowdManagedBounty} or Hackerone \cite{hackeroneHackerOne}) -- sometimes referred to as bug bounty platforms -- 
provide a seemingly convenient way to act as an intermediary
between those who discover a vulnerability and those who are affected.
Walshe et al.~\cite{Walshe.2020} conducted a cost-benefit analysis of bug bounty programs for companies. They found that these programs provide ``a valuable complementary technique'' for a company's cybersecurity management, as vulnerabilities are found at a decent rate and the costs for bug bounty programs are usually lower than hiring software engineers~\cite{Walshe.2020}.

However, when validating the quality of reports within such bug bounty programs, Shafigh et al.~\cite{Shafigh.2021} found that many reports marked invalid were either labeled ``out-of-scope'' or ``false positive'', meaning the scope of what qualifies as a bug for a company was probably not entirely clear to the reporters. 
By analyzing CVD policy documents of companies on bug bounty platforms, Walshe et al.~\cite{Walshe.2023} found that CVD policy documents contain a large number of legal references and only little relevant information, e.g., on out-of-scope reports or legal consequences, which would make it a requirement for those reporting a vulnerability to have a full understanding of the laws and provisions in place while providing less relevant information on the reporting process itself.

From an organization's perspective, reports are often too numerous to be manageable, especially when companies lack formalized methods to process such reports \cite{Smale.2023} or when the reports contain too little actionable information \cite{Wunder.2024}. 
In a retrospective study, Walshe et al.~\cite{Walshe.2022} identified several fears prior to launching a CVD program (e.g., distrust in those disclosing vulnerabilities), 
various issues after launching (e.g., concerning the quality and scope of reported vulnerabilities), and 
internal 
issues
, e.g., issues with the quality and scope of the reported vulnerabilities, poor communication with those reporting the vulnerabilities, or challenges in identifying responsible product owners~\cite{Walshe.2022}. 

Walshe et al.~\cite{Walshe.2022} recruited companies from bug bounty platforms on their experiences with CVD programs, meaning the companies had already a CVD program in place at the time of the survey. We are not aware of related studies that have surveyed a representative sample of companies, including those with no or little experience with CVD programs. This motivated our second research question: \textit{\textbf{RQ2:} \rqtwo}

\section{Methodology}
\label{sec:methodology}
\begin{figure*}[t]
    \centering
    \includegraphics[width=0.85\linewidth]{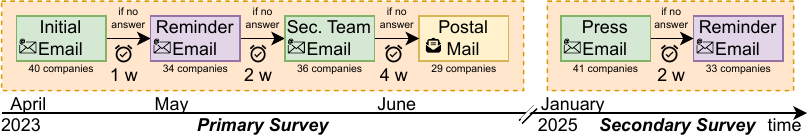}
    \caption{Timeline and contact phases of the \primarysurvey~and \secondarysurvey.}
    \label{fig:survey-phases}
\end{figure*}

To answer our first research question, we examined the existence of publicly available information about CVD programs and security.txt files. 
To answer RQ2, we collected contact information for various points of contact within the 40 German DAX companies in our sample to invite them to a survey and analyze their answers.
The \primarysurvey~was conducted in April and May 2023, and a \secondarysurvey~to better understand developments was carried out in January 2025 (see \Cref{fig:survey-phases}). 

\subsection{Company Sample}
\rev{We selected leading German companies as the basis for our sample to learn about the status quo of CVD programs and changes in their adoption.} 
We placed our focus on Germany, as it \rev{is} the economic leader of the EU \cite{statistaEuropesBiggest}. Therefore, its leading companies should be at the forefront of innovation, cybersecurity, and technological best practices, serving as a role model for SMEs in Germany and companies from other countries. Furthermore, Germany was still in the process of implementing the new EU provisions at that time, \rev{so we could use this 
transition phase to learn about their effect on the motivations and challenges of implementing CVD processes.}

We deliberately chose Germany's 40 largest and most important publicly traded companies tracked by the DAX\footnote{``Deutscher Aktienindex'', English: German stock index} stock market index (see \Cref{sec:appendix:dax-companies}) due to their economic importance in the country, their coverage of multiple critical industries (e.g., automotive, finance, healthcare, technology), and their large number of worldwide employees and customers. We assume that these companies have to handle large amounts of personal or sensitive data and manage numerous IT systems, thus requiring proper IT security practices, including the ability to respond to external security reports. 
Additionally, shareholders and stakeholders may have expectations of proper IT security and data protection, with CVD processes as one aspect.

Our final sample included the 40 DAX companies (2023 - 2024), plus one additional company in the \secondarysurvey~due to a change in the DAX stock index as of January 1, 2025.

\subsection{Public CVD Information}
\label{sec:presurvey:meth:publiccvd}
To answer RQ1, we primarily focused on publicly available information about the companies' CVD programs. \rev{We collected and analyzed the available information once between April 30, 2023, and May 15, 2023, during the \primarysurvey{} and a second time on January 24, 2025, at the end of the \secondarysurvey{}.} 
First, we checked for security.txt files \cite{rfc9116} on the companies' websites (see \Cref{tab:dax-companies} in \Cref{sec:appendix:dax-companies}) in accordance with the paths and protocol defined in the RFC~\cite{rfc9116}.
Second, we used a search engine (Google.com) or browsed a company's website to look for references to a public coordinated vulnerability disclosure program, e.g., by searching for ``\emph{<company name> vulnerability disclosure}''\footnote{\rev{The English search term ``<company> vulnerability disclosure'' was chosen to find pages related to ``coordinated vulnerability disclosure'' (CVD), ``vulnerability disclosure program'' (VDP), or other pages containing the keywords ``vulnerability'' and ``disclosure'' assumed to be present in bug bounty policies or responsible disclosure programs, as these terms appear to be adapted into German language.}}

\subsection{CVD Survey}
\subsubsection{Contact Channels}
\label{sec:presurvey:meth:contactinfo}
In order to increase the chances of receiving a reply to our survey, we used different contact addresses to approach the companies (see \Cref{fig:survey-phases}) via electronic and postal mail\footnote{Companies are required to provide such contact information under German law \cite{tmg_5,ddg_5}.} in the \primarysurvey. For the \secondarysurvey, we used only email.

Contact information was collected manually from the companies' websites. We searched for ``\emph{<company name> Impressum}'' (English: ``Imprint'' or legal notice) to find the respective legal page if we could not locate it by navigating the website. If we could not identify any company-related email address, we used the respective contact forms on the website. 
Additionally, we tried to find contact information for a company's security team. One established way to publish such contact information is through a \emph{security.txt} file, as described in RFC 9116~\cite{rfc9116}. If we could not identify an email address in a security.txt file, we navigated the website and used a Google's search engine to discover IT security-related contact information. If none could be found, we used the RFC 2142-recommended functional mailbox \emph{security@<domain.tld>} \cite{rfc2142}.
For postal mail, we used the company's headquarters address provided on the legal notice (``Imprint'') page.

For the \secondarysurvey, we decided to approach the press departments, as we received only a few responses and reactions to our \primarysurvey{} (see \Cref{sec:discussion:lessons-learned} for a detailed explanation). We deliberately chose not to contact the security teams (again) since we could not identify such contact information for all companies, and many security contacts explicitly stated they should be contacted only for reporting vulnerabilities.
We considered press teams to be more appropriate contacts, as they usually manage a company's public relations and would be able to forward our request. Similar to the \primarysurvey, we navigated the websites to find pages titled
``press'' or ``media'', or used a search engine with the keyword ``\emph{<company name> Pressekontakt}'' (English: press contact) to find the respective contact email addresses or forms. 
Some companies did not list a general email address but individual email addresses to press team members with dedicated areas of expertise (e.g., information security), which we then used instead of general contact addresses.

\subsubsection{Contact Phases}

The \primarysurvey~was conducted in April and May 2023 in four consecutive contact attempts as depicted in \Cref{fig:survey-phases}. The first three attempts focused on email, while the last phase used traditional postal mail. Companies that replied to our inquiry with survey answers or explicitly declined participation were excluded from the remaining phases of the same survey. In the \secondarysurvey~we contacted all companies at most twice and only by email.

\begin{itemize}
    \item \primarysurvey{} (S1)
    \begin{enumerate}
        \item \emph{Initial Email:}
        We sent the first batch of questionnaires via email and contact forms to the contacts collected from the legal notice pages on April 23, 2023. 
        \item \emph{Reminder Email:}
        One week later, we sent reminder emails to the contacts who had not replied. 
        \item \emph{Security Team Email:}
        As the questionnaire primarily consisted of IT security-related questions, we also decided to send an email to the company's security team if we did not receive an answer two weeks after sending a reminder email.
        \item \emph{Postal Mail:}
        After a month without answers in any previous phase, we sent a letter to their postal addresses. 
        We hoped to avoid automatic spam filters or other technical barriers while increasing the likelihood that a human would read our inquiry. In addition to the cover letter and the questionnaire, our letter included a prepaid and labeled return envelope for the company's convenience.
    \end{enumerate}
    \item \secondarysurvey{} (S2)
    \begin{enumerate}
        \item \emph{Press Email:}
        For the \secondarysurvey, we sent out the first batch of questionnaires via email and contact forms to the contacts collected from the press and media pages of all 41 companies on January 10, 2025.
        \item \emph{Reminder Email:}
        Two weeks later, we sent a reminder email to the same contacts if we had not received an answer. We extended the time between the initial email and reminder compared to the \primarysurvey~since we received some out-of-office responses to our initial email, and, further, learned from the \primarysurvey~that some companies need more time to answer. 
    \end{enumerate}
\end{itemize}

The surveyed companies were asked to respond directly to our questions via electronic or postal mail. 
 We deliberately chose not to use external tools (e.g., web-based survey platforms) to reduce the risk of interference by security scanners or spam filters, and the risk of being mistaken for a phishing attempt \cite{fette.2007,toolan.2010}. Furthermore, typing links from letters is inconvenient.
 
\subsubsection{Questionnaire and Cover Letter}
In order to answer RQ2, we sent out questionnaires between April and May 2023, as well as in January 2025.
Our cover letter included an introductory text that explained our motivation for the survey and our affiliation with the university. We also included our contact information (address, email, phone) for any questions from the contacted companies and explained that the answers would be analyzed in an aggregated manner to preserve anonymity. Finally, the companies were asked to answer our questions of their own volition as a response to our (e)mail. For the postal mail, we also included a pre-stamped return envelope to minimize the effort required to respond. 

We designed a questionnaire with 13 questions attached to the cover letter (see~\Cref{sec:appendix:questionnaire-initial}) for the \primarysurvey, to analyze the existence and implementation of CVD programs, and the company's experiences with them. 
We provided possible answer categories for our questions to give respondents an indication of the type of response we expected. However, each question could also be answered completely openly.

For the \secondarysurvey~we slightly rephrased the introductory text and added a very brief explanation
of CVD programs, clearly stated our research goal, and mentioned that we had already contacted the company on this topic between April and May 2023. Furthermore, we also added an English translation of the cover letter and the questionnaire, as some companies had only one press team for the entire international corporation, and we could not be entirely sure that all recipients spoke German.
We shortened the questionnaire to the five most relevant questions in the \secondarysurvey~(see~\Cref{sec:appendix:questionnaire-retention}), as we assumed that the large number of questions in the \primarysurvey~might have deterred some companies from responding. Another rationale was that a shortened questionnaire would take less time to complete and might reduce the impression that too much ``sensitive information'' is being given away while still providing us with enough insights into their CVD adoption.
The shortened questionnaire focused solely on the existence and details of a CVD program or insights into the \rev{absence} of one. Additionally, we asked the companies about their challenges, benefits, and drawbacks of CVD programs. 
The new questions were very similar (if not identical) to the longer questionnaire, allowing us to map and merge the answers (see \Cref{sec:appendix:answers}).

\begin{table*}[t]
	\centering
	\caption{Distribution of the questionnaire answers to our CVD survey from the German 40 DAX companies.
	}
	\label{tab:responses}
	\begin{tabularx}{\textwidth}{l *{6}{X}*{1}{l}}
		\toprule
		& \multicolumn{4}{c}{\textbf{\primarysurvey}} & \multicolumn{2}{c}{\textbf{\secondarysurvey}} & \multicolumn{1}{c}{\textbf{\textit{Total}}}\\
		\cmidrule(lr){2-5} \cmidrule(lr){6-7} \cmidrule(lr){8-8}
		& Initial Email & Reminder Email & Sec-Team Email & Postal Mail & Press Email & Reminder Email \\
		\midrule
		\textbf{Contacted Companies} & 40 & 34 & 36 & 29 & 41 & 33 & -\\
		\textbf{Responses Received} & 6 & 0 & 5 &  1 & 8 & 11 & 31\\
		\textbf{Questionnaires Answered} & 0 & 0 & 4 & 1 & 1 & 2 & 8\\
		\bottomrule
	\end{tabularx}
\end{table*}

\section{\rev{Results}}
\label{sec:results}
\rev{This section presents the results from the publicly collected information on CVD programs and security contacts, which we used to answer \textbf{RQ1}, and the results from surveying the 40 DAX companies, which helped us to answer \textbf{RQ2}.}

To obtain our results, we tracked all relevant information in a single spreadsheet during both surveys and used several approaches to analyze our data. We used simple descriptive analysis to report the adoption rate of public CVD programs and security.txt files (see~\Cref{sec:results:cvdprograms}). We aggregated the answers from both questionnaires (see \Cref{sec:appendix:answers}) to provide an overview of all responses. 
Since we primarily used closed questions with possible answers provided (e.g., ``Yes / No / uncertain'' or ``high costs / low or no added value / other''), we mainly used quantitative methods to analyze the responses (see~\Cref{sec:results:survey}). Due to the low number of responses, it was not reasonable to perform any statistical testing, so we relied solely on descriptive analysis.

\subsection{RQ1: Status Quo \& Adoption of CVD Programs}
\label{sec:results:cvdprograms}
\begin{figure}
    \centering
    \includegraphics[width=\linewidth]{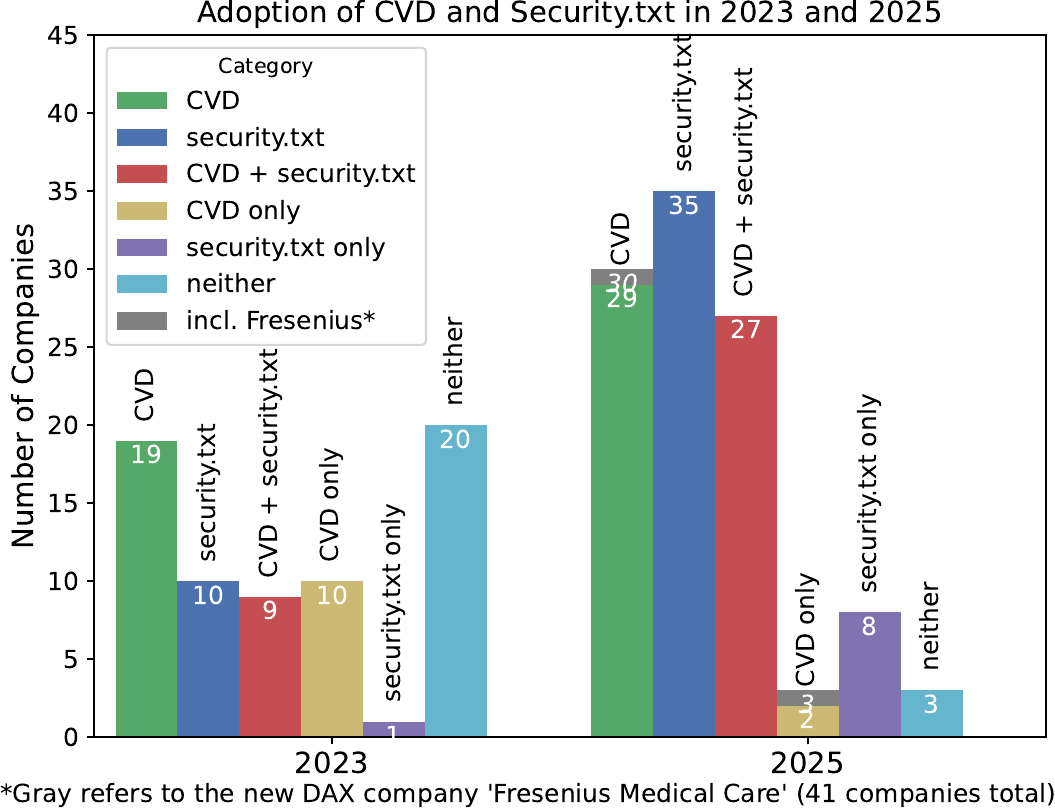}
    \caption{Comparison of the adoption of CVD programs and security.txt files in 2023 vs. 2025.}
    \label{fig:cvd-comparison-2023-2025}
\end{figure}
As visualized in \Cref{fig:cvd-comparison-2023-2025}, \rev{our analysis} showed that only 50\% (20/40) of the German DAX companies provided opportunities to disclose vulnerabilities via security.txt files or CVD programs in 2023. 
Of those, 47.5\% (19/40) had established a public CVD process.
Nine of these 19 also had a security.txt file. Only one company had a security.txt file but no CVD program, while ten companies had a CVD program but no security.txt file. In total, one company provided only a security.txt file, nine companies provided both, and ten companies had only a CVD program in place. 
This means we could not identify a security-related contact channel during the \primarysurvey{} for 20 companies.

During our second analysis in January 2025, we observed a substantial increase in the ability to disclose vulnerabilities to 92.5\% (37/40). 
\rev{At the time of writing,} almost three-quarters (29/40 = 72.5\%) of the originally analyzed companies had established a CVD program to accept vulnerability reports compared to 2023, while 87.5\% (35/40) have a security.txt file to provide a point of contact, which is important \rev{according to Kranenbarg et al. \cite{Kranenbarg.2018}}. 
Including the new DAX company that joined the index on January 1, 2025, the number of CVD programs increased from 19 to 30, and the number of security.txt deployments has increased from ten to 35, as 25 companies started using a security.txt file between 2023 and 2025.
While 27 companies (67.5\%) use both security mechanisms, three companies (7.5\%) have not yet provided vulnerability disclosure possibilities in the form of a security.txt file or CVD program.
For the 20 companies where we were unable to find a security.txt file or CVD program in 2023, the analysis after the \secondarysurvey~shows the following: ten companies now use a security.txt file \emph{and} a CVD program, and seven companies deployed a security.txt file, but no CVD program.

In total, we identified ten new CVD programs and 25 new security.txt files, increasing vulnerability disclosure contact possibilities to 92.5\% (37/40) for the initial set of DAX companies compared to 2023.
The high adoption rate should be viewed as a positive development from a security perspective, as it indicates that companies are becoming more receptive to external security reports.

While we recognized that several companies established their CVD program or security.txt files in 2023 or throughout 2024 -- several months after our \primarysurvey{} -- we have no proof that this was a result of our messages. However, it is possible that our study helped raise awareness about this topic as a side effect. 
Raising awareness for the existence of CVD programs and the implementation of security contacts is important for those reporting vulnerabilities \cite{Kranenbarg.2018}. The security.txt file, which must be deployed at specific paths and formatted in a structured way, is a sensible solution to guide those who want to report a security vulnerability. In fact, many of the security.txt files in our analysis contained an email address or a link to the CVD program, ensuring that the security team is the first point of contact for security researchers.

\subsection{RQ2: Challenges \& Experiences with CVD Programs}
\label{sec:results:survey}

We received survey responses from eight different companies (five for the \primarysurvey{} (S1, {[P1 -- P5]}) and three for the \secondarysurvey{} (S2, {[P6 -- P8]}) as shown in \Cref{tab:responses}) that we analyzed to answer \textbf{RQ2}. All responses were distinct, as no company answered both surveys.

Due to the low response rate of only 20\%, we aggregated the answers from both surveys for the analysis. We analyzed most of the answers \rev{using descriptive statistics}, as the low number of total responses did not allow for meaningful statistical testing. For the few open-ended questions, \rev{we used a simplified qualitative approach in which} two researchers reviewed all responses and grouped similar answers \rev{into broad} categories. This was done in a joint process.
There was no need to conduct a thorough thematic analysis or develop a codebook, as all answers were very brief and could be easily grouped (e.g., S1 [Q9], S2 [Q3]: \sq{``more security''}, \sq{``You gain more security''}, \sq{``improved security''}, \sq{``security''}, \sq{``security +''}, and \sq{``security +''} were grouped as \sq{``more security (6x)''}, see \Cref{tab:appendix:isanswers} and \Cref{tab:appendix:fsanswers} in Appendix \ref{sec:appendix:answers}).

Concerning the existence of CVD programs (S1 [Q1, Q2], S2 [Q1]), the answers showed that three companies run a CVD program on an external CVD platform, five others run their own CVD program; two use both approaches, and one is in the process of launching a CVD program. One company launched its program more than 11 years ago [P3], another in 2020 [P1], one was launched at the beginning of 2023 [P2], and one during 2024 [P8], with another in the process of being established [P6]. The companies with an operational CVD program in place also provided the following details about their programs: within the last year, they received between 11-100 (2x), or up to 1,000 security reports (S1 [Q5]). For most companies, the reports were mostly valid (80 -- 100\%: 3x; 60 -- 80\%: 1x), but one reported that only 20 -- 40\% of the reports were valid (S1 [Q5a]). Two rated the reports' quality as good, one very good, and one \sq{``mixed''} (S1 [Q5b]). 
Vulnerability reports are prioritized based on the severity or business impact, individual case-by-case decisions, or standardized CVSSv3 scores (S1 [Q7]). Two companies noted that the vulnerability types vary greatly (S1 [Q13]), and valid issues are usually remediated within two months (S1 [Q8]: 10-30 days (2x) or 60+ days (1x)). One company has a CVD budget of 10k - 100k EUR, another over 1m EUR, and others did not answer this question (S1 [Q6]).
The effectiveness of their CVD programs (S1 [Q12]) is measured based on the number of submissions, the quality of findings, the ratio between valid and invalid submissions or their severity, and the number of uncoordinated vulnerability disclosures reaching zero.

As benefits from implementing CVD programs (S1 [Q9], S2 [Q3]), the companies mentioned increasing the company's IT security (6x, e.g., [P8]: \sq{``Improved security''}) or increasing trust and reputation (3x, e.g., [P1]: \sq{``Trust from partners and customers''}, [P4]: \sq{``Reputation +''}). Two participants stated legal compliance as a motivating factor ([P3]: \sq{``legal compliance''}, [P6]: \sq{``The primary goal is to comply with legal requirements (NIS2, EU CRA).''}). Further benefits identified by companies include preventing attacks ([P1]: \sq{``better self-protection''}), facilitating structured input channels ([P2]: \sq{``Organized input channel (conciseness)''}), as well as adopting a better approach to deal with mistakes ([P7]: \sq{``good approach to dealing with mistakes [...] A good approach to dealing with mistakes also benefits the company's image. In addition, despite a competent, active red team, there are still things here and there that have remained undetected. So you gain more security.''}) and acquiring competence ([P3]: \sq{``Gaining expertise within the company''}).

With respect to challenges in implementing CVD processes (S1 [Q3], S2 [Q2]), companies described a lack of resources (human or financial, e.g., [P1]: \sq{``lack of resources, [...] costs''}), lack of security awareness ([P3]: \sq{``lower security awareness at that point''}), as well as \sq{``quality \& quantity of findings''} [P1], and \sq{``raising hackers' interest due to launching the program''} [P2]. [P4] also raised the issue of \sq{``unclear responsibilities of the individual departments.''}

Again, a lack of resources in terms of budget or personnel (e.g., [P1]: \sq{``high effort and costs''}, [P2]: \sq{''staff costs''}) was also mentioned as a hurdle for running CVD programs (S1 [Q10, Q11], S2 [Q4, Q5]). Further issues include \sq{``lack of responsibilities''} [P4], or difficulties in finding qualified staff ([P1]: \sq{``qualified personnel hard to find''}). In fact, from the \primarysurvey{}, only three organizations have dedicated personnel to handle the CVD program; one does not, and one preferred not to answer (S1 [Q4]).
Interestingly, two companies stated that they do not see any major issues in running CVD programs -- especially when compared to other business processes. [P7] answered: \sq{``Hurdles [are] not that big. You need a budget for bounties every year and it must be clear that you also need resources for community management and that this cannot be done ``on the side''. The program needs a structured concept, which is reviewed by the company's lawyers and data protection officers, and then you're ready to go. However, this is the case for every business process [...], so there are hardly any differences in terms of the hurdles.''} [P8] added to that: \sq{``Wouldn't say it was a big challenge, but setting up the process and a policy consumes some time.''}

Related to challenges in cooperation with other stakeholders (S1 [Q11], S2 [Q5]), companies mentioned integration of and discussions with external service providers or customers, as well as identifying all relevant parties (e.g., [P7]: \sq{``It may happen that external service providers, who, for example, provide a specialized IT function, explicitly wish to be excluded from consideration by third parties within the scope of bug bounty programs. They would like us to include a long list of exceptions in our bug bounty program.''}, or [P2]: \sq{``Finding proper contact persons''}). Further challenges included \sq{``high workload''} [P3], or \sq{``budget''} [P4].

While most of the companies eventually implemented CVD processes, [P6] can serve as an interesting case study, as they were still in the process of implementing it. Notably, while a CVD program is still being developed, they have already deployed a security.txt file: \sq{``In the short term, security.txt was established. The introduction of a CVD process is currently being implemented.''} As described above, the main benefits of a CVD process for [P6] are structuring vulnerability reporting, and ensuring legal compliance: \sq{``The primary goal is to comply with legal requirements (NIS2, EU CRA). In addition, coordination is important. Channeling input and providing a corresponding description of how it is handled facilitates processing.''} 
They stated that internal coordination processes and integration into existing processes are their main challenges: \sq{``Coordinating with the departments involved across the entire company is a challenge. The process must be integrated into the existing process landscape (quality management).''} The challenges the company faces with respect to external stakeholders include integrating customers into the vulnerability management process and managing vulnerabilities in general. They aim for a Common Security Advisory Framework (CSAF) as an ideal communication method, but describe this as \sq{``a long way to go''} (\sq{``We see customer integration as a challenge. Addressing vulnerabilities is one thing. Informing customers with the appropriate contacts is difficult. In the best-case scenario, communication is automated (see CSAF). There is still a long way to go.''}). However, so far the company does not see any disadvantages in CVD programs. 

While we received answers to our questionnaires from only 20\% (8/40) of the surveyed companies, as discussed in \Cref{sec:recommendations:limitations}, 
they helped us better understand the challenges and experiences associated with CVD programs. Similar to Walshe et al.~\cite{Walshe.2022}, companies reported challenges such as missing (human) resources combined with high workloads or large numbers of invalid reports. However, the majority of the companies reported that the reports they receive are of (very) good quality. It appears that misconceptions about CVD programs, particularly regarding the expected quality of reports, may differ from reality. Future work should investigate this in more detail.

In contrast, most of the answers suggest that CVD programs are overall a valuable addition to the companies' security, as also found by Walshe et al.~\cite{Walshe.2020}. Overall, the survey responses indicate that the benefits of running a CVD program outweigh the effort required. In fact, one company even indicated that establishing a CVD program is not too much of a hurdle, as it is not much different from establishing other business processes that require dedicated time and resources. We observed that several surveyed companies have implemented or plan to implement CVD programs to enhance their security. While intrinsic motivation, i.e., protecting their own company or increasing trust and reputation, was mentioned several times, it is interesting to note that legal compliance is also mentioned as a motivator by two companies. Thus, raising awareness and strengthening intrinsic motivation are important factors. However, creating legal certainty by transposing EU directives (e.g., CRA, NIS2) into national law as soon as possible appears to be the most effective motivator for fostering the implementation of vulnerability disclosure processes. 

Most interestingly, one company provided reasons for not having implemented a CVD program so far. While they described internal obstacles (e.g., not yet having a vulnerability management process in place), they also highlighted the need for improvements to the Common Security Advisory Framework (CSAF) to communicate with external stakeholders -- an aspect that was also investigated by Wunder et al.~\cite{Wunder.2024}. 

However, legal compliance was the reason for this company to at least deploy a security.txt file as a standardized first point of contact for vulnerability disclosure. This aspect could be a valuable lesson for other companies that are still hesitant or do not yet have the resources to implement a proper CVD program. Especially in the context of upcoming laws, adopting initial CVD practices (e.g., security.txt files) early and later extending the program's scope cannot only help overcome potential challenges or concerns but also foster better security within organizations. 

\section{\rev{Discussion}}
\label{sec:discussion}
\rev{This section primarily discusses the results in light of the lessons learned from surveying the German DAX companies and how these insights can benefit SMEs or policymakers, before addressing ethical considerations, limitations, and directions for future work.}

\subsection{Lessons Learned} 
\label{sec:discussion:lessons-learned}

\Cref{tab:responses} provides an overview of the number of contacted companies, the number of responses, and the number of answers to the questionnaire we received throughout the different contact phases of our \primarysurvey{} and \secondarysurvey{}. In total, we received 31 responses to our survey emails, resulting in a 20\% (8/40) answer rate to our survey questionnaire, which helped us better understand the challenges and experiences associated with CVD programs. 

During the \primarysurvey{}'s \emph{first contact phase} (``Initial Email''), we only received two positive responses acknowledging the receipt of the survey and stating that survey answers would be provided later, but no responses to our questionnaire were given. Four declined to answer the questionnaire, and the rest did not respond to our contact attempts.
In the \textit{second phase} (``Reminder Email''), we received no responses to our survey at all.
Within the \emph{third phase} (``Sec-Team Email''), we included two companies that acknowledged our inquiry but did not provide any answers. We received four responses with answers to our questionnaire {([P1] -- [P4])}, which, interestingly, all came from the ten manually collected security team addresses, while 46\% (12/26) of the generic \emph{security@<domain.tld>} email addresses resulted in delivery errors (bounces). The fifth response stated that the security team was not allowed to answer such questionnaires.
In the \textit{fourth phase} (``Postal Mail''), the letters, prepared with considerable effort and time, were sent to the companies via postal mail only resulted in just one additional completed questionnaire {([P5])}. We did not receive any other reaction (i.e., via email) to our postal letters. 
Thus, we received a total of five unique questionnaire answers in the \primarysurvey, which equals an overall response rate of 12.5\% (5/40). 

Within the \secondarysurvey, we received eight replies to our first email (``Press Email'') on January 10, 2025, but only one completed our questionnaire {([P6])}. Note that this was a company that had not answered our \primarysurvey{} before. Of the remaining responses, one company answered that they would forward the inquiry, one indicated that these questions were too specialized and could, thus, not be answered, two declined to participate in university- or research-related surveys (e.g., due to high demand), and three declined to answer our inquiry (e.g., due to too many surveys or inquiries).

Exactly two weeks later, we sent a reminder to all companies (``Reminder Email'') that had not answered in the first phase of the \secondarysurvey{} yet. This time, the reminder led to a higher response rate: a total of 11 companies replied, but seven declined to participate in the survey for various reasons (i.e., lack of capacity, no participation in surveys, lack of university support, or inability to answer our questions). One company requested that we forward the survey to a specific employee, and another forwarded our request internally; however, we did not receive a completed questionnaire in either case. Notably, two other DAX companies replied with answers to our questionnaire later, both of which had not answered in the previous phases of the \primarysurvey{} {([P7, P8])}.

Through our multi-layered contact-phase approach, we learned that contacting the same company through different channels can yield better results. In some cases, we received an answer after initially receiving a refusal to participate from a different point of contact, or the other way around: we received an answer in the \primarysurvey{}, but a refusal in the \secondarysurvey{}.
In contrast to our expectations, contacting companies by postal mail did not yield the desired effect of more survey answers -- we received only one. 
Using only email would also make it easier to include links and other resources, such as online surveys, references, or attachments, which could have a positive impact on anonymization or response rates.

We found that using the generic \textit{security@<domain.tld>} address resulted in delivery failures in almost 50\% of the cases, indicating that these companies do not follow the recommendations of RFC 2142 \cite{rfc2142}. Low delivery rates for generic emails \rev{have} also been reported in previous studies~\cite{Stock.2016, Utz.2023, Stock.2018}, making this type of contact information very unreliable compared to collecting contact information directly from websites~\cite{Maass.2021, Maass.2021w1o, Hennig-Dietmann.2022, Hennig2022, Utz.2023}.
However, when a dedicated security email address was provided, we were able to reach suitable points of contact successfully. Interestingly, we received most of the survey answers directly from the security teams. However, we discovered during the \secondarysurvey{} that many security teams state that their contact addresses are only to be used for vulnerability reports.
Thus, we recommend that future work carefully assess whether a security team can be contacted for non-vulnerability matters.

Overall, we learned that reaching out to multiple points of contact via email with reminders after several weeks yielded the best chances of obtaining answers to our questionnaire. 

\subsection{Recommendations}
\label{sec:discussion:recommendations}

\subsubsection{CVD Adoption}
Our results show that most DAX companies now have a security.txt file or CVD program in place to facilitate the coordinated reporting of vulnerabilities. 
Although various challenges, such as high effort and costs, were noted, many answers indicate a positive effect on their company's security, e.g., through high-quality or diverse types of vulnerabilities being reported.
While SMEs may not have equally high budgets available for vulnerability rewards, dedicated personnel, or CVD platforms, hosting a security.txt file or a reward-free CVD program could be a first step in aligning with upcoming EU rules and offering security researchers a less frustrating reporting experience. 
Furthermore, openness to external security reports can help mitigate data breaches or other security incidents while providing legal guarantees for ethical hackers. 
The observed trend of increasing CVD programs in DAX companies can serve as a positive signal to SMEs to adopt similar strategies. 
Future work should monitor adoption within SMEs and conduct similar surveys to learn about the differences in challenges and experiences faced by SMEs.

\subsubsection{CVD Regulation}
Our longitudinal analysis falls within the timeframe of new security provisions in the EU, which encourage the establishment of CVD programs. 
We observed the creation of new security.txt files and CVD programs between 2023 and 2025; however, we cannot determine whether this was due to our inquiries or other factors.
However, the answers did indicate that some of these programs were launched in anticipation of the upcoming NIS2 and CRA rules.

Nonetheless, during the collection of contact information, we noticed that although all companies have published contact information for their data protection officers in accordance with GDPR requirements, finding security contact information is not as straightforward (as also reported by Poteat et al.~\cite{Poteat.2021}).
Some companies host security.txt files with or without contact email addresses, while others only provide contact forms or place security contact information elsewhere on their website.
Therefore, we recommend that policymakers discuss and establish similar contact detail requirements for security reports as for data protection inquiries (e.g., a dedicated ``security officer'' and contact page).

\subsection{Ethical Considerations}
\label{sec:recommendations:ethics}
According to the ethics board\footnote{
\ifisanonymized
The link will be added in the camera-ready version.
\else
\url{https://ethikkommission.eecs.tu-berlin.de/}
\fi
} of the first author's institution, approval for this study was not required, as it did not meet any of the four conditions that would necessitate review (e.g., participants exposed to risk or belonging to a vulnerable group, experiencing physical or emotional stress, or not being fully informed about the study). Nevertheless, we completed the university's self-assessment form, which addresses these criteria to the best of our knowledge, and did not identify any aspects of concern.

\newpage
In a few cases, companies were contacted again despite having previously declined in an earlier phase of the \primarysurvey, which might raise ethical concerns. Although unintentional and likely due to the short response window in the \primarysurvey{}, we took steps to minimize repetition in the \secondarysurvey{} by extending the time we waited for answers before sending the reminder. We did not receive any complaints throughout our study.

Since we promised the companies to handle their data confidentially, we feel obliged to keep the original datasets private. However, we believe that the described methodology and the public nature of most of the information allow future work to reproduce our study with limited effort.

\ifisanonymized
\emph{Note to reviewers: We provide you the anonymized spreadsheets and answers used for our results in the following repository:
\url{https://anonymous.4open.science/r/German-DAX-CVD-research-F8C6/}}
\fi

\subsection{Limitations}
\label{sec:recommendations:limitations}
We utilized our university's infrastructure for sending emails, which is configured to state-of-the-art standards (e.g., SPF, DMARC, DKIM), according to the IT department. However, network issues, spam filters, or other factors could still have influenced the successful delivery, nonetheless. To mitigate email delivery issues, we sent reminders and postal mail in the \primarysurvey.

Our manual public CVD program analysis is limited by how Google's search algorithm indexes websites, ranks search results, and processes the chosen search keywords, as we only considered the first page of results. 
To mitigate this limitation, we compared our results with the community-driven project ``disclose.io''\footnote{\url{https://disclose.io/programs/}}, but found it contained fewer CVD programs than we identified.

Since the periods between phases in the \primarysurvey{} were relatively short and companies needed more time to respond than anticipated, there may have been overlaps between the time at which the companies replied to our inquiry from a previous phase and when we sent new inquiries for the next phase. We are aware of only a few cases in which companies that declined to answer in a previous phase were nonetheless included in subsequent phases. Also, during the fourth phase of the \primarysurvey{}, two companies with similar names were inadvertently omitted due to human error.

Furthermore, due to the data protection policy of the first and second author's university, we unexpectedly lost access to the email account used for correspondence during the \primarysurvey{}. According to the university's IT department, restoring access to the account or the emails was not possible. However, we are confident that the (anonymized) data transferred to the evaluation spreadsheet is correct (with respect to the survey answers). Nevertheless, the limited meta information does not allow us to determine exactly in which phase the rejections were received. 
Hence, in contrast to what we had planned for the methodology, it is possible that companies that had already declined were contacted again. As we did not receive any negative feedback from the companies, we are confident that this did not negatively affect our results.

The generalizability of our results is limited by the small sample size (only 40 companies from one country) and by the modest number of questionnaire answers we received from the companies (eight in total), which further limited the generalizability of our results and made our survey more of a ``case study''.
While the sample is limited, focusing on these top companies allowed us to test several contact channels (e.g., security teams or press departments), which might not exist for SMEs. Furthermore, contacting a larger number of SMEs was deemed infeasible for this study due to the high manual effort to identify the necessary contact information, leaving this task for future work. 
Nonetheless, we are confident that the insights into these companies' challenges and experiences provide a valuable source for deriving recommendations on implementing CVD processes in SMEs or discussing new legal changes with policymakers. 
Furthermore, we believe that we can contribute valuable lessons learned regarding contact channels and starting points for future work, e.g., for comparisons with other EU countries.

\subsection{Future Work}
\label{sec:future-work}
There are several aspects we consider interesting for future work. First, both surveys had a low response rate despite using several different contact channels. We observed a high bounce rate for generic \emph{security@<domain.tld>} email addresses, consistent with findings in related work \cite{Stock.2016,Stock.2018,Utz.2023}, making this the least effective notification channel. While we considered general contact information or press contacts the most appropriate\footnote{Especially when a company has a public CVD program in place, we expected the press team to answer at least our first question about its existence, but we received many refusals to participate instead.} for our inquiry, reaching out to security contacts, i.e., email addresses provided in the security.txt file, yielded the highest number of answers. However, many companies do not provide such information (yet), and, furthermore, these points of contact should primarily be used for vulnerability reports, as also stated in multiple CVD program policies. 
For reporting vulnerabilities, however, these results are encouraging: it seems that if a dedicated contact is provided, it can be successfully reached.

Thus, it might also be worthwhile reaching out to data protection officers (DPOs), similar to the work of Hennig et al.~\cite{Hennig-Dietmann.2022}, since some vulnerabilities can also lead to privacy incidents, making vulnerability disclosure processes an issue for DPOs as well. For companies processing personal data, the appointment of a data protection officer is mandatory, as per Article 37 of the GDPR. In the future, national laws implementing NIS2 may follow a similar approach and mandate a designated ``security officer'' as the first point of contact for vulnerability reports or security incidents. 

Future work could also investigate reasons for non-responses or identify appropriate contact channels in more detail, as, for example, some press contacts declined our inquiry because it was deemed out of scope. For example, in contrast to the results of Maass et al~\cite{Maass.2021w1o}, sending surveys via postal mail did not improve response rates but involved high costs and effort. Other options include reaching out to companies via social networks (e.g., LinkedIn) or by phone, or becoming a shareholder in the case of publicly traded companies. One would then be in a position to ask questions, for example, about (missing) CVD processes and their effect on the company's stock valuation, at the annual shareholder meetings.

Furthermore, a longitudinal study monitoring the adoption of security.txt files or CVD programs with a larger sample, including SMEs or international companies, could reveal more detailed information on how security processes evolve. Finally, this could provide interesting insights and further results on why companies do not implement or have not yet implemented CVD processes.  

\section{Summary \& Conclusion}
\label{sec:conclusion}
This paper analyzed the landscape of Coordinated Vulnerability Disclosure programs among German DAX companies. We observed significant progress in the adoption of publicly available CVD programs and security.txt files, reaching up to 92.5\% between 2023 and 2025, leaving only three companies without security contact information. 

Although we received many answers from the DAX companies to our contact attempts via electronic or postal mail, only 20\% of them completed our questionnaire. However, these answers provided valuable insights into the introduction of and experiences with CVD programs, bridging the gap in the literature between companies that have CVD practices implemented and those that do not, and offering valuable insights for policymakers and other companies. For example, we learned that legal compliance (e.g., NIS2, CRA) or improving the company's security are strong motivators for adopting CVD programs; however, companies also express concerns about a lack of human resources and missing internal processes.

By asking companies about their CVD processes, we may have helped raise awareness of CVD programs; however, a broader survey could reveal more general insights.
Future work should explore methods to increase response rates, as we did not receive answers from all companies that we contacted. 

Overall, the positive trend of newly introduced CVD programs and security.txt files over the past two years promises to make it easier for individuals to disclose vulnerabilities in the future.

\section*{Acknowledgment}
\addcontentsline{toc}{section}{Acknowledgment}
The third author's contributions to this work were supported by funding from the project ``Engineering Secure Systems'' of the Helmholtz Association (HGF) [topic 46.23.01] and by KASTEL Security Research Lab. Special thanks to Julian Hunter for his private lesson in European legislation.


%
%
%
\newpage
\bibliographystyle{IEEEtran}
\bibliography{euroUsec}

@misc{ddg_5,
    author = {{Bundesamt für Justiz}}, 
    title  = {DDG - Digitale-Dienste-Gesetz *},
    url = {https://www.gesetze-im-internet.de/ddg/BJNR0950B0024.html},
    date = {2025-05-6},
    note = {[Accessed 07-07-2025]}

}

@misc{tmg_5,
	author = {dejure.org},
	title = {§ 5 {T}{M}{G} - {A}llgemeine {I}nformationspflichten - dejure.org --- dejure.org},
	url = {https://dejure.org/gesetze/TMG/5.html},
	year = {2021},
    date = {2021-08-10},
	note = {[Accessed 12-01-2025]},
}

@misc{rfc9116,
	author = {Shafranovich, Y. and Foudil, E.},
	title = {{R}{F}{C} 9116: {A} {F}ile {F}ormat to {A}id in {S}ecurity {V}ulnerability {D}isclosure --- datatracker.ietf.org},
	url = {https://datatracker.ietf.org/doc/html/rfc9116},
	year = {2022},
	note = {[Accessed 12-01-2025]},
}

@misc{rfc2142,
	author = {Dave Crocker},
	title = {{R}{F}{C} 2142: {M}ailbox {N}ames for {C}ommon {S}ervices, {R}oles and {F}unctions --- datatracker.ietf.org},
	url = {https://datatracker.ietf.org/doc/html/rfc2142},
	year = {1997},
	note = {[Accessed 12-01-2025]},
}

@misc{haveibeenpwnedHaveBeen,
	author = {Troy Hunt},
	title = {{H}ave {I} {B}een {P}wned: {C}heck if your email has been compromised in a data breach --- haveibeenpwned.com},
	url = {https://haveibeenpwned.com/},
	year = {2025},
	note = {[Accessed 07-03-2025]},
}

@misc{hackeroneHackerOne,
	author = {HackerOne},
	title = {{H}acker{O}ne --- hackerone.com},
	url = {https://hackerone.com/hacktivity/overview},
	year = {2025},
	note = {[Accessed 07-03-2025]},
}

@misc{bugcrowdManagedBounty,
	author = {Bugcrowd},
	title = {{M}anaged bug bounty programs, a better approach to security --- bugcrowd.com},
	url = {https://bugcrowd.com/crowdstream},
	year = {2025},
	note = {[Accessed 07-03-2025]},
}

@misc{statistaEuropesBiggest,
	author = {Aaron O'Neill},
	title = {{G}{D}{P} of {E}urope's biggest economies 1980-2029 | {S}tatista --- statista.com},
	url = {https://www.statista.com/statistics/959301/gdp-of-europes-biggest-economies/},
	year = {2025},
    date = {10.01.2025},
	note = {[Accessed 07-03-2025]},
}

@article{Stoever.2023,
  title={{How website owners face privacy issues: Thematic analysis of responses from a covert notification study reveals diverse circumstances and challenges}},
  author={St{\"o}ver, Alina and Gerber, Nina and Prid{\"o}hl, Henning and Maass, Max and Bretthauer, Sebastian and Hollick, Matthias and Herrmann, Dominik and others},
  journal={Proceedings on Privacy Enhancing Technologies},
  year={2023},
  doi={10.56553/popets-2023-0051}
}

@inproceedings{Utz.2023,
           title = {{Comparing Large-Scale Privacy and Security Notifications}},
            year = {2023},
          author = {Christine Utz and Matthias Michels and Martin Degeling and Ninja Marnau and Ben Stock},
           month = {July},
         journal = {Proceedings on Privacy Enhancing Technologies},
       booktitle = {PETS 2023},
             url = {https://publications.cispa.saarland/3918/},
}

@misc{Hove.2023,
      title={{Your Vulnerability Disclosure Is Important To Us: An Analysis of Coordinated Vulnerability Disclosure Responses Using a Real Security Issue}}, 
      author={Koen van Hove and Jeroen van der Ham-de Vos and Roland van Rijswijk-Deij},
      year={2023},
      eprint={2312.07284},
      archivePrefix={arXiv},
      primaryClass={cs.NI},
      url={https://arxiv.org/abs/2312.07284}, 
        note = {[Accessed 20-01-2025]},
}

@INPROCEEDINGS{Lone.2022,
  author={Lone, Qasim and Frik, Alisa and Luckie, Matthew and Korczyński, Maciej and van Eeten, Michel and Gañán, Carlos},
  booktitle={{2022 IEEE Symposium on Security and Privacy (SP)}}, 
  title={{Deployment of Source Address Validation by Network Operators: A Randomized Control Trial}}, 
  year={2022},
  volume={},
  number={},
  pages={2361-2378},
  doi={10.1109/SP46214.2022.9833701},
  url={https://ieeexplore.ieee.org/abstract/document/9833701}
}

@inproceedings{Poteat.2021,
author = {Poteat, Tara and Li, Frank},
title = {{Who you gonna call? an empirical evaluation of website security.txt deployment}},
year = {2021},
isbn = {9781450391290},
publisher = {Association for Computing Machinery},
address = {New York, NY, USA},
url = {https://doi.org/10.1145/3487552.3487841},
doi = {10.1145/3487552.3487841},
booktitle = {{Proceedings of the 21st ACM Internet Measurement Conference}},
pages = {526–532},
numpages = {7},
location = {Virtual Event},
series = {IMC '21}
}

@inproceedings{Wunder.2024,
author = {Wunder, Julia and Aurich, Janik and Benenson, Zinaida},
title = {{From Chaos to Consistency: The Role of CSAF in Streamlining Security Advisories}},
year = {2024},
isbn = {9798400717963},
publisher = {Association for Computing Machinery},
address = {New York, NY, USA},
url = {https://doi.org/10.1145/3688459.3688463},
doi = {10.1145/3688459.3688463},
booktitle = {{Proceedings of the 2024 European Symposium on Usable Security}},
pages = {187–199},
numpages = {13},
location = {
},
series = {EuroUSEC '24}
}

@inproceedings{Findlay.2022,
  title={{Characterizing the adoption of security. txt files and their applications to vulnerability notification}},
  author={Findlay, W Paul and Abdou, Abdelrahman},
  booktitle={{Proceedings of the Workshop on Measurements, Attacks, and Defenses for the Web (MADWeb)}},
url = {https://www.ndss-symposium.org/ndss-paper/auto-draft-282/},
  year={2022}
}

@inproceedings {Maass.2021,
author = {Max Maass and Alina St{\"o}ver and Henning Prid{\"o}hl and Sebastian Bretthauer and Dominik Herrmann and Matthias Hollick and Indra Spiecker},
title = {{Effective Notification Campaigns on the Web: A Matter of Trust, Framing, and Support}},
booktitle = {{30th USENIX Security Symposium (USENIX Security 21)}},
year = {2021},
isbn = {978-1-939133-24-3},
pages = {2489--2506},
url = {https://www.usenix.org/conference/usenixsecurity21/presentation/maass},
publisher = {USENIX Association},
month = aug
}

@article{Çetin.2016,
    author = {Çetin, Orçun and Hanif Jhaveri, Mohammad and Gañán, Carlos and van Eeten, Michel and Moore, Tyler},
    title = {{Understanding the role of sender reputation in abuse reporting and cleanup}},
    journal = {Journal of Cybersecurity},
    volume = {2},
    number = {1},
    pages = {83-98},
    year = {2016},
    month = {12},
    issn = {2057-2085},
    doi = {10.1093/cybsec/tyw005},
    url = {https://doi.org/10.1093/cybsec/tyw005},
    eprint = {https://academic.oup.com/cybersecurity/article-pdf/2/1/83/10833175/tyw005.pdf},
}

@inproceedings{Durumeric.2014,
author = {Durumeric, Zakir and Li, Frank and Kasten, James and Amann, Johanna and Beekman, Jethro and Payer, Mathias and Weaver, Nicolas and Adrian, David and Paxson, Vern and Bailey, Michael and Halderman, J. Alex},
title = {{The Matter of Heartbleed}},
year = {2014},
isbn = {9781450332132},
publisher = {Association for Computing Machinery},
address = {New York, NY, USA},
url = {https://doi.org/10.1145/2663716.2663755},
doi = {10.1145/2663716.2663755},
booktitle = {{Proceedings of the 2014 Conference on Internet Measurement Conference}},
pages = {475–488},
numpages = {14},
location = {Vancouver, BC, Canada},
series = {IMC '14}
}

@inproceedings{Li.2016,
author = {Li, Frank and Ho, Grant and Kuan, Eric and Niu, Yuan and Ballard, Lucas and Thomas, Kurt and Bursztein, Elie and Paxson, Vern},
title = {{Remedying Web Hijacking: Notification Effectiveness and Webmaster Comprehension}},
year = {2016},
isbn = {9781450341431},
publisher = {International World Wide Web Conferences Steering Committee},
address = {Republic and Canton of Geneva, CHE},
url = {https://doi.org/10.1145/2872427.2883039},
doi = {10.1145/2872427.2883039},
booktitle = {{Proceedings of the 25th International Conference on World Wide Web}},
pages = {1009–1019},
numpages = {11},
location = {Montr\'{e}al, Qu\'{e}bec, Canada},
series = {WWW '16}
}

@inproceedings {Stock.2016,
	author = {Ben Stock and Giancarlo Pellegrino and Christian Rossow and Martin Johns and Michael Backes},
	title = {{Hey, You Have a Problem: On the Feasibility of {Large-Scale} Web Vulnerability Notification}},
	booktitle = {{25th USENIX Security Symposium (USENIX Security 16)}},
	year = {2016},
	isbn = {978-1-931971-32-4},
	address = {Austin, TX},
	pages = {1015--1032},
	url = {https://www.usenix.org/conference/usenixsecurity16/technical-sessions/presentation/stock},
	publisher = {USENIX Association},
	month = aug
}

@inproceedings {Vasek.2012,
	title = {{Do Malware Reports Expedite Cleanup? An Experimental Study}},
	booktitle = {{5th Workshop on Cyber Security Experimentation and Test (CSET 12)}},
	year = {2012},
	address = {Bellevue, WA},
	url = {https://www.usenix.org/conference/cset12/workshop-program/presentation/Vasek},
	publisher = {USENIX Association},
	month = aug
}

@inproceedings{Maass.2021w1o,
author = {Maa\ss{}, Max and Clement, Marc-Pascal and Hollick, Matthias},
title = {{Snail Mail Beats Email Any Day:On Effective Operator Security Notifications in the Internet}},
year = {2021},
isbn = {9781450390514},
publisher = {Association for Computing Machinery},
address = {New York, NY, USA},
url = {https://doi.org/10.1145/3465481.3465743},
doi = {10.1145/3465481.3465743},
booktitle = {{Proceedings of the 16th International Conference on Availability, Reliability and Security}},
articleno = {11},
numpages = {13},
location = {Vienna, Austria},
series = {ARES '21}
}

@inproceedings {Çetin.2018,
author = {Or{\c c}un {\c C}etin and Carlos Ga{\~n}{\'a}n and Lisette Altena and Samaneh Tajalizadehkhoob and Michel van Eeten},
title = {{Let Me Out! Evaluating the Effectiveness of Quarantining Compromised Users in Walled Gardens}},
booktitle = {{Fourteenth Symposium on Usable Privacy and Security (SOUPS 2018)}},
year = {2018},
isbn = {978-1-939133-10-6},
address = {Baltimore, MD},
pages = {251--263},
url = {https://www.usenix.org/conference/soups2018/presentation/cetin},
publisher = {USENIX Association},
month = aug
}

@inproceedings {Li.20167ep,
	author = {Frank Li and Zakir Durumeric and Jakub Czyz and Mohammad Karami and Michael Bailey and Damon McCoy and Stefan Savage and Vern Paxson},
	title = {{You{\textquoteright}ve Got Vulnerability: Exploring Effective Vulnerability Notifications}},
	booktitle = {{25th USENIX Security Symposium (USENIX Security 16)}},
	year = {2016},
	isbn = {978-1-931971-32-4},
	address = {Austin, TX},
	pages = {1033--1050},
	url = {https://www.usenix.org/conference/usenixsecurity16/technical-sessions/presentation/li},
	publisher = {USENIX Association},
	month = aug
}

@article{Stock.2018,
author = "Ben Stock and Giancarlo Pellegrino and Frank Li and Michael Backes and Christian Rossow",
title = {{Didn't You Hear Me? --- Towards More Successful Web Vulnerability Notifications}},
year = "2018",
month = "2",
url = "https://publications.cispa.de/articles/conference_contribution/Didn_t_You_Hear_Me_---_Towards_More_Successful_Web_Vulnerability_Notifications/24612648",
doi = "10.60882/cispa.24612648.v1"
}

@inproceedings{Zeng.2019,
  title={{Fixing HTTPS misconfigurations at scale: An experiment with security notifications}},
  author={Zeng, Eric and Li, Frank and Stark, Emily and Felt, Adrienne Porter and Tabriz, Parisa},
  booktitle={{Workshop on the Economics of Information Security}},
  year={2019},
  url={https://d1wqtxts1xzle7.cloudfront.net/99658198/li-weis2019-libre.pdf}
}

@inproceedings{Çetin.2017,
  title={{Make notifications great again: learning how to notify in the age of large-scale vulnerability scanning}},
  author={Cetin, Orcun and Ganan, Carlos and Korczynski, Maciej and Van Eeten, Michel},
  booktitle={{Workshop on the Economics of Information Security (WEIS)}},
  volume={23},
  year={2017},
url = {https://repository.tudelft.nl/record/uuid:621f4a4f-e5d9-4f04-abc4-46252f9db3db}
}

@INPROCEEDINGS{Çetin.2019,
  author={Çetin, Orçun and Gañán, Carlos and Altena, Lisette and Tajalizadehkhoob, Samaneh and van Eeten, Michel},
  booktitle={{2019 IEEE European Symposium on Security and Privacy (EuroS\&P)}}, 
  title={{Tell Me You Fixed It: Evaluating Vulnerability Notifications via Quarantine Networks}}, 
  year={2019},
  volume={},
  number={},
  pages={326-339},
url = {https://ieeexplore.ieee.org/abstract/document/8806733},
  doi={10.1109/EuroSP.2019.00032}}

@article{StopBadware.2012, 
year = {2012}, 
title = {{Compromised Websites: An Owner's Perspective}}, 
author = {{StopBadware and Commtouch}}, 
url = {https://www.stopbadware.org/files/compromised-websites-an-owners-perspective.pdf}, 
pages = {1 -- 15}
}

@techreport{CMU.2017, 
year = {2017}, 
author = {Householder, Allen D. and Wassermann, Garrett and Manion, Art and King, Chris}, 
title = {{The CERT Guide to Coordinated Vulnerability Disclosure}}, 
url = {https://insights.sei.cmu.edu/documents/1945/2017\_003\_001\_503340.pdf}, 
urldate = {2025-01-21}
}

@techreport{ENISA.2022, 
year = {2022}, 
author = {{European Union Agency for Cybersecurity (ENISA)}}, 
editorb = {Evangelos Kantas,Marnix Dekker}, 
editora = {Débora Di Giacomo,Nick Conway (Wavestone),Aude Thirriot (Wavestone),Thiago Barbizan (Wavestone),Solène Drugeot (Wavestone),Cristian Michael Tracci (Wavestone),Lorenzo Pupillo (Centre for European Policy Studies (CEPS)),Carolina Polito (CEPS),Francesco Campoli}, 
title = {{Coordinated Vulnerability Disclosure policies in the EU - Coordinated Vulnerability Disclosure policies in the EU}}, 
url = {https://www.enisa.europa.eu/sites/default/files/publications/Coordinated Vulnerability Disclosure policies in the EU.pdf}, 
urldate = {2025-01-21}
}

@techreport{BSI.2022, 
year = {2022}, 
author = {{German Federal Office for Information Security}}, 
title = {{Leitlinie des BSI zum Coordinated Vulnerability Disclosure (CVD)-Prozess}}, 
url = {https://www.bsi.bund.de/SharedDocs/Downloads/DE/BSI/CVD/CVD-Leitlinie.pdf?\_\_blob=publicationFile\&v=4}, 
urldate = {2025-01-21}
}

@misc{Iso.2018, 
year = {2018}, 
title = {{ISO/IEC 29147:2018(en), Information technology — Security techniques — Vulnerability disclosure}}, 
author = {{ISO Central Secretary}}, 
url = {https://www.iso.org/obp/ui/en/\#iso:std:iso-iec:29147:ed-2:v1:en}, 
urldate = {2025-01-21}
}

@article{Kranenbarg.2018, 
year = {2018}, 
title = {{Don’t shoot the messenger! A criminological and computer science perspective on coordinated vulnerability disclosure}}, 
author = {Kranenbarg, Marleen Weulen and Holt, Thomas J. and Ham, Jeroen van der}, 
journal = {Crime Science}, 
doi = {10.1186/s40163-018-0090-8}, 
url = {https://link.springer.com/article/10.1186/s40163-018-0090-8#citeas},
pages = {16}, 
number = {1}, 
volume = {7}
}

@Inbook{Pil.2023,
author="Pil, Yoon Sang",
editor="Lee, Roger",
title={{The Way Forward for Security Vulnerability Disclosure Policy: Comparative Analysis of US, EU, and Netherlands}},
bookTitle={{Big Data, Cloud Computing, and Data Science Engineering}},
year="2023",
publisher="Springer International Publishing",
address="Cham",
pages="119--131",
isbn="978-3-031-19608-9",
doi="10.1007/978-3-031-19608-9_10",
url="https://doi.org/10.1007/978-3-031-19608-9_10"
}

@INPROCEEDINGS{Walshe.2020,
  author={Walshe, Thomas and Simpson, Andrew},
  booktitle={{2020 IEEE 2nd International Workshop on Intelligent Bug Fixing (IBF)}}, 
  title={{An Empirical Study of Bug Bounty Programs}}, 
  year={2020},
  volume={},
  number={},
  pages={35-44},
  url= {https://ieeexplore.ieee.org/abstract/document/9034828},
  doi={10.1109/IBF50092.2020.9034828}}

@article{Vostoupal.2024,
title = {{The legal aspects of cybersecurity vulnerability disclosure: To the NIS 2 and beyond}},
journal = {Computer Law \& Security Review},
volume = {53},
pages = {105988},
year = {2024},
issn = {2212-473X},
doi = {https://doi.org/10.1016/j.clsr.2024.105988},
url = {https://www.sciencedirect.com/science/article/pii/S0267364924000554},
author = {Jakub Vostoupal and Václav Stupka and Jakub Harašta and František Kasl and Pavel Loutocký and Kamil Malinka},
}

@misc{Ruohonen.2024,
      title={{Vulnerability Coordination Under the Cyber Resilience Act}}, 
      author={Jukka Ruohonen and Paul Timmers},
      year={2025},
      eprint={2412.06261},
      archivePrefix={arXiv},
      primaryClass={cs.CR},
      url={https://arxiv.org/abs/2412.06261},  
	note = {[Accessed 07-03-2025]}
}

@misc{NIS2.2022, 
title={{D}irective - 2022/2555 of the {E}uropean {P}arliament and of the {C}ouncil of 14 {D}ecember 2022 on measures for a high common level of cybersecurity across the {U}nion}, 
url={https://eur-lex.europa.eu/eli/dir/2022/2555/oj/eng}, 
author={{European Union}}, 
year={2022},
note = {[Accessed 07-03-2025]}
}

@inproceedings {Kuehrer.2014,
	author = {Marc K{\"u}hrer and Thomas Hupperich and Christian Rossow and Thorsten Holz},
	title = {{Exit from Hell? Reducing the Impact of {Amplification} {DDoS} Attacks}},
	booktitle = {{23rd USENIX Security Symposium (USENIX Security 14)}},
	year = {2014},
	isbn = {978-1-931971-15-7},
	address = {San Diego, CA},
	pages = {111--125},
	url = {https://www.usenix.org/conference/usenixsecurity14/technical-sessions/presentation/kuhrer},
	publisher = {USENIX Association},
	month = aug
}

@INPROCEEDINGS{Nosyk.2023,
  author={Nosyk, Yevheniya and Korczyński, Maciej and Gañán, Carlos H. and Król, Michał and Lone, Qasim and Duda, Andrzej},
  booktitle={{2023 IEEE 22nd International Conference on Trust, Security and Privacy in Computing and Communications (TrustCom)}}, 
  title={{Don’t Get Hijacked: Prevalence, Mitigation, and Impact of Non-Secure DNS Dynamic Updates}}, 
  year={2023},
  volume={},
  number={},
  pages={1480-1489},
url = {https://ieeexplore.ieee.org/abstract/document/10538543},
  doi={10.1109/TrustCom60117.2023.00202}}

@InProceedings{Hennig-Dietmann.2022,
author="Hennig, Anne
and Dietmann, Heike
and Lehr, Franz
and Mutter, Miriam
and Volkamer, Melanie
and Mayer, Peter",
editor="Clarke, Nathan
and Furnell, Steven",
title={{``Your Cookie Disclaimer is Not in Line with the Ideas of the GDPR. Why?''}},
booktitle={{Human Aspects of Information Security and Assurance}},
year="2022",
publisher="Springer International Publishing",
address="Cham",
pages="218--227",
isbn="978-3-031-12172-2",
url={https://link.springer.com/chapter/10.1007/978-3-031-12172-2_17}
}

@article{Rodriguez.2021,
    author = {Rodríguez, Elsa and Verstegen, Susanne and Noroozian, Arman and Inoue, Daisuke and Kasama, Takahiro and van Eeten, Michel and Gañán, Carlos H},
    title = {{User compliance and remediation success after IoT malware notifications}},
    journal = {{Journal of Cybersecurity}},
    volume = {7},
    number = {1},
    pages = {tyab015},
    year = {2021},
    month = {07},
    issn = {2057-2085},
    doi = {10.1093/cybsec/tyab015},
    url = {https://doi.org/10.1093/cybsec/tyab015},
    eprint = {https://academic.oup.com/cybersecurity/article-pdf/7/1/tyab015/39047609/tyab015.pdf},
}

@article{Cetin.2019ndss,
title={{Cleaning Up the Internet of Evil Things: Real-World Evidence on ISP and Consumer Efforts to Remove Mirai}},
author={Orçun Çetin and Carlos Hernandez Ga{\~n}{\'a}n and Lisette Altena and Takahiro Kasama and Daisuke Inoue and Kazuki Tamiya and Ying Tie and Katsunari Yoshioka and Michel van Eeten},
journal={{Proceedings 2019 Network and Distributed System Security Symposium}},
year={2019},
url={https://doi.org/10.14722/ndss.2019.23438}
}

@inproceedings{Shafigh.2021,
author = {Shafigh, Saman and Benatallah, Boualem and Rodr\'{\i}guez, Carlos and Al-Banna, Mortada},
title = {{Why Some Bug-bounty Vulnerability Reports are Invalid? Study of bug-bounty reports and developing an out-of-scope taxonomy model}},
year = {2021},
isbn = {9781450386654},
publisher = {Association for Computing Machinery},
address = {New York, NY, USA},
url = {https://doi.org/10.1145/3475716.3484193},
doi = {10.1145/3475716.3484193},
booktitle = {{Proceedings of the 15th ACM / IEEE International Symposium on Empirical Software Engineering and Measurement (ESEM)}},
articleno = {38},
numpages = {6},
location = {Bari, Italy},
series = {ESEM '21}
}

@article{Schmitz.2021, 
year = {2021}, 
title = {{Responsible Vulnerability Disclosure under the NIS 2.0 Proposal}}, 
author = {Schmitz, Sandra and Schiffner, Stefan}, 
journal = {{Journal of Intellectual Property, Information, Technology, and Electronic Commerce Law}}, 
url = {https://www.jipitec.eu/jipitec/article/view/336/}, 
pages = {448--457}, 
number = {12}, 
volume = {5},
}

@INPROCEEDINGS{Smale.2023,
  author={de Smale, Stephanie and van Dijk, Rik and Bouwman, Xander and van der Ham, Jeroen and van Eeten, Michel},
  booktitle={{2023 IEEE Symposium on Security and Privacy (SP)}}, 
  title={{No One Drinks From the Firehose: How Organizations Filter and Prioritize Vulnerability Information}}, 
  year={2023},
  volume={},
  number={},
  pages={1980-1996},
url = {https://ieeexplore.ieee.org/abstract/document/10179447},
  doi={10.1109/SP46215.2023.10179447}}

@article{Moura.2023,
author = {Moura, Giovane C. M. and Heidemann, John},
title = {{Vulnerability Disclosure Considered Stressful}},
year = {2023},
issue_date = {April 2023},
publisher = {Association for Computing Machinery},
address = {New York, NY, USA},
volume = {53},
number = {2},
issn = {0146-4833},
url = {https://doi.org/10.1145/3610381.3610383},
doi = {10.1145/3610381.3610383},
journal = {{SIGCOMM Comput. Commun. Rev.}},
month = jul,
pages = {2–10},
numpages = {9},
url= {https://dl.acm.org/doi/abs/10.1145/3610381.3610383}
}

@article{Walshe.2023,
author = {Walshe, Thomas and Simpson, Andrew},
title = {{Towards a Greater Understanding of Coordinated Vulnerability Disclosure Policy Documents}},
year = {2023},
issue_date = {June 2023},
publisher = {Association for Computing Machinery},
address = {New York, NY, USA},
volume = {4},
number = {2},
url = {https://doi.org/10.1145/3586180},
doi = {10.1145/3586180},
journal = {Digital Threats},
month = aug,
articleno = {29},
numpages = {36}
}

@article{Vandezande.2024,
title = {{Cybersecurity in the EU: How the NIS2-directive stacks up against its predecessor}},
journal = {Computer Law \& Security Review},
volume = {52},
pages = {105890},
year = {2024},
issn = {2212-473X},
doi = {https://doi.org/10.1016/j.clsr.2023.105890},
url = {https://www.sciencedirect.com/science/article/pii/S0267364923001000},
author = {Niels Vandezande},
}

@article{Walshe.2022,
title = {{Coordinated Vulnerability Disclosure programme effectiveness: Issues and recommendations}},
journal = {Computers \& Security},
volume = {123},
pages = {102936},
year = {2022},
issn = {0167-4048},
doi = {https://doi.org/10.1016/j.cose.2022.102936},
url = {https://www.sciencedirect.com/science/article/pii/S0167404822003285},
author = {T. Walshe and A.C. Simpson}
}

@INPROCEEDINGS{Chen.2024,
  author={Chen, Ting-Han and Tagliaro, Carlotta and Lindorfer, Martina and Borgolte, Kevin and Van Der Ham-De Vos, Jeroen},
  booktitle={{2024 IEEE European Symposium on Security and Privacy Workshops (EuroS\&PW)}}, 
  title={{Are You Sure You Want To Do Coordinated Vulnerability Disclosure?}}, 
  year={2024},
  volume={},
  number={},
  pages={307-314},
url = {https://ieeexplore.ieee.org/abstract/document/10628550},
  doi={10.1109/EuroSPW61312.2024.00039}}

@INPROCEEDINGS{toolan.2010,
  author={Toolan, Fergus and Carthy, Joe},
  booktitle={{2010 eCrime Researchers Summit}}, 
  title={{Feature selection for Spam and Phishing detection}}, 
  year={2010},
  volume={},
  number={},
  pages={1-12},
  doi={10.1109/ecrime.2010.5706696}}

@inproceedings{fette.2007,
author = {Fette, Ian and Sadeh, Norman and Tomasic, Anthony},
title = {Learning to detect phishing emails},
year = {2007},
isbn = {9781595936547},
publisher = {Association for Computing Machinery},
address = {New York, NY, USA},
url = {https://doi.org/10.1145/1242572.1242660},
doi = {10.1145/1242572.1242660},
booktitle = {Proceedings of the 16th International Conference on World Wide Web},
pages = {649–656},
numpages = {8},
location = {Banff, Alberta, Canada},
series = {WWW '07}
}

@article{hilbig.2023,
author = {Hilbig, Tobias and Geras, Thomas and Kupris, Erwin and Schreck, Thomas},
title = {security.txt Revisited: Analysis of Prevalence and Conformity in 2022},
year = {2023},
issue_date = {September 2023},
publisher = {Association for Computing Machinery},
address = {New York, NY, USA},
volume = {4},
number = {3},
url = {https://doi.org/10.1145/3609234},
doi = {10.1145/3609234},
journal = {Digital Threats},
month = oct,
articleno = {36},
numpages = {17}
}

@article{Hennig2022, 
  year    = {2022}, 
  title   = {Standing out among the daily spam: How to catch website owners’ attention by means of vulnerability notifications}, 
  author  = {Hennig, Anne and Neusser, Fabian and Pawelek, Aleksandra Alicja and Herrmann, Dominik and Mayer, Peter}, 
  journal = {{CHI} Conference on Human Factors in Computing Systems Extended Abstracts}, 
  doi     = {10.1145/3491101.3519847}, 
  pages   = {1--8}
}
%

\clearpage
\newpage
\appendix
\appendices

\section{German DAX Companies}\label{sec:appendix:dax-companies}
\begin{table}[ht]
\centering
\caption{German DAX companies and their websites}
\label{tab:dax-companies}
\small
\begin{adjustbox}{max width=\linewidth}
\begin{tabular}{@{}ll@{}}
\toprule
\textbf{Company Name} & \textbf{Website} \\
\midrule
adidas AG & \url{www.adidas.de} \\
Airbus SE & \url{www.airbus.com} \\
Allianz SE & \url{www.allianz.de} \\
BASF SE & \url{www.basf.com} \\
Bayer AG & \url{www.bayer.com} \\
Beiersdorf AG & \url{www.beiersdorf.de} \\
BMW AG & \url{www.bmw.de} \\
Brenntag SE & \url{www.brenntag.com} \\
Commerzbank AG & \url{www.commerzbank.de} \\
Continental AG & \url{www.continental.com} \\
Covestro AG$^\dagger$ & \url{www.covestro.com} \\
Daimler Truck Holding AG & \url{www.daimlertruck.com} \\
Deutsche Bank AG & \url{www.deutsche-bank.de} \\
Deutsche Börse AG & \url{www.deutsche-boerse.com} \\
Deutsche Telekom AG & \url{www.telekom.de} \\
DHL Group (ex Deutsche Post AG) & \url{www.deutschepost.de} \\
E.ON SE & \url{www.eon.de} \\
Fresenius SE \& Co. KGaA & \url{www.fresenius.com} \\
Hannover Rück SE & \url{www.hannover-re.com} \\
Heidelberg Material AG & \url{www.heidelbergmaterials.de} \\
Henkdel AG \& Co. KGaA & \url{www.henkel.de} \\
Infineon AG & \url{www.infineon.com} \\
Mercedes-Benz Group AG & \url{www.mercedes-benz.com} \\
MERCK KGaA & \url{www.merckgroup.com} \\
MTU Aero Engines AG & \url{www.mtu.de} \\
Müchener Rückversichungs-Gesellschaft AG & \url{www.munichre.com} \\
Porsche AG & \url{www.porsche.com} \\
Porsche Automobil Holding SE & \url{www.porsche-se.com} \\
QIAGEN NV & \url{www.qiagen.com} \\
Rheinmetall AG & \url{www.rheinmetall.com} \\
RWE AG & \url{www.rwe.com} \\
SAP SE & \url{www.sap.com} \\
Sartorius AG VZ & \url{www.sartorius.com} \\
Siemens AG & \url{www.siemens.com} \\
Siemens Energy & \url{www.siemens-energy.com} \\
Siemens Healthineers AG & \url{www.siemens-healthineers.com} \\
Symrise AG & \url{www.symrise.com} \\
Volkswagen AG VZ & \url{www.volkswagenag.de} \\
Vonovia SE & \url{www.vonovia.de} \\
Zalando SE & \url{www.zalando.com} \\
(Fresenius Medical Care St.)$^\oplus$ & \url{www.freseniusmedicalcare.com} \\
\bottomrule
\end{tabular}
\end{adjustbox}
$^\dagger${Dropped out 2024-12-31}, $^\oplus${Joined 2025-01-01}
\end{table}

\section{\rev{Survey Questionnaires}}\label{sec:appendix:questionnaire}
\subsection{\primarysurvey~(S1)}\label{sec:appendix:questionnaire-initial}
\begin{enumerate}
    \item [Q1)] Are you enrolled in a Bug Bounty program or using an external CVD platform? (Yes / No)
    \item [Q2)] Have you established a CVD program within your organization? (Yes / No)
    \begin{enumerate}
        \item If yes, how long has your CVD program been in place, and why was it introduced? (e.g., response to past attacks / prevention / inspiration from other companies)
        \item If no, do you plan to establish a CVD program in the future? (Yes / No / Uncertain)
        \item  If no or uncertain, what is holding you back? (e.g., specific concerns / obstacles)
    \end{enumerate}
    \item [Q3)] What are the main problems your organization faces in implementing a CVD program, or what were the challenges prior to its introduction? (e.g., lack of resources / lack of expertise / legal concerns / other)
    \item [Q4)] Does your organization have specifically trained or assigned personnel for conducting CVD activities? (Yes / No / Uncertain)
    \begin{enumerate}
        \item If no, who is responsible, or which department handles CVD-related tasks? (e.g., IT security officer / developer / other)
    \end{enumerate}
    \item [Q5)] How many vulnerabilities have been reported to your organization by external security researchers in the last 12 months? (0-10 / 11-100 / 101-1,000 / 1,000+)
    \begin{enumerate}
        \item What percentage of the reported vulnerabilities are valid? (0-20\% / 20-40\% / 40-60\% / 60-80\% / 80-100\%)
        \item How would you rate the quality of the vulnerability reports? (very poor / poor / good / very good)
    \end{enumerate}
    \item [Q6)] What is the budget allocated to your organization for conducting CVD activities? (No budget / Less than 10,000 / 10,000-100,000 / 100,000-1,000,000 / 1,000,000+ EUR)
    \item [Q7)] How does your organization triage and prioritize reported vulnerabilities? (e.g., based on severity / based on business impact / other)
    \item [Q8)] How long does it typically take for your organization to fix a reported vulnerability? (Less than 10 days / 10-30 days / 30-60 days / 60+ days)
    \item [Q9)] What are the main benefits of a CVD program for your organization? (e.g., improved security / increased trust with customers and partners / other)
    \item [Q10)] What are the main drawbacks of a CVD program for your organization? (e.g., high costs / low or no added value / other)
    \item [Q11)] What are the most important challenges in collaborating with other stakeholders in the CVD process? (e.g., lack of communication / lack of trust / legal concerns / other)
    \item [Q12)] How do you measure the effectiveness of your CVD program? (e.g., number of resolved vulnerabilities/number of researchers reporting vulnerabilities / other)
    \item [Q13)] What types of vulnerabilities have been predominantly uncovered so far? (Cross Site Scripting / Authorization / Information / Sensitive Data Exposure / Business Logic / Content / Authentication / Server Security / SQL Injection / CSRF / Remote Code / DoS / Brute Force / Other)
\end{enumerate}

\subsection{\secondarysurvey~(S2)}\label{sec:appendix:questionnaire-retention}
\begin{enumerate}
    \item [Q1)] Has your organization created mechanisms to report security vulnerabilities through a CVD program (e.g., security.txt [2], self-managed or externally managed programs like CVD platforms such as HackerOne, Bugcrowd, or Intigriti)? (Yes / No)
    \begin{enumerate}
        \item If yes:
        \begin{enumerate}
            \item What measures or types of CVD programs do you use?
            \item When was your CVD program established?
            \item Why was it introduced (e.g., response to past attacks / prevention / inspiration from other companies / legal requirements)? 
        \end{enumerate}
        \item If no, do you plan to establish a CVD program in the future? (Yes / No / Uncertain) 
        \item If no or uncertain, what prevents you from doing so (e.g., specific concerns / obstacles)?
    \end{enumerate}
\end{enumerate}
If your organization participates in a Bug Bounty program or has established a CVD program:
\begin{enumerate}
\setcounter{enumi}{1}
    \item [Q2)] What were the biggest challenges for your company before introducing a CVD program (e.g., lack of resources / lack of expertise / legal concerns / other)?
    \item [Q3)] What are the main benefits of a CVD program for your organization (e.g., improved security / increased trust with customers and partners / other)?
    \item [Q4)] What are the main drawbacks of a CVD program for your organization (e.g., high costs / low or no added value / other)?
    \item [Q5)] What are the key challenges in collaborating with other stakeholders in the CVD process (e.g., lack of communication / lack of trust / legal concerns / other)?
\end{enumerate}

\section{\rev{Survey Answers}}\label{sec:appendix:answers}
\Cref{tab:appendix:isanswers} and \Cref{tab:appendix:fsanswers} 
contain the anonymized and aggregated answers to our \primarysurvey{} (S1) and \secondarysurvey{} (S2).
The answers can be merged and aggregated using the following mapping, as the questions are similar: 
\begin{multicols}{2}
\begin{itemize}[left=0pt]
\small
    \item S2 (Q1) $\rightarrow$ S1 (Q2)
    \item S2 (Q1a I) $\rightarrow$ S1 (Q2)
    \item S2 (Q1a II) $\rightarrow$ S1 (Q2a)
    \item S2 (Q1a III) $\rightarrow$ S1 (Q2a)
    \item S2 (Q1b) $\rightarrow$ S1 (Q2b)
    \item S2 (Q1c) $\rightarrow$ S1 (Q2c)
    \item S2 (Q2) $\rightarrow$ S1 (Q3)
    \item S2 (Q3) $\rightarrow$ S1 (Q9)
    \item S2 (Q4) $\rightarrow$ S1 (Q10)
    \item S2 (Q5) $\rightarrow$ S1 (Q11)
\end{itemize}
\end{multicols}

\begin{table}[!h]
	\centering
	\caption{Anonymized and aggregated answers we received during the \secondarysurvey.}
	\label{tab:appendix:fsanswers}
	\resizebox{\linewidth}{!}{%
		\begin{tabular}{@{}p{0.13\linewidth}p{0.9\linewidth}@{}}
			\toprule
			\textbf{Question}    & \textbf{Answers}                                            \\ \midrule
			(Q1)         & Yes (3x)                                           \\
			(Q1a I)   & security.txt (3x), CVD program in progress, self-managed  (2x), bug-bounty on CVD platform (1x)         \\
			(Q1a II)  & 02.2024, 2024                                            \\
			(Q1a III) & prevention (2x), legal obligations, professional handling of vulnerabilities                    \\
			(Q1b)     & -                                                  \\
			(Q1c)     & -                                                  \\ \midrule
			(Q2) & coordination with all internal departments, process has to be integrated in existing quality management, budget \& resources, a structured concept checked by legal and data protection departments, setting up processes and a policy takes time \\ \midrule
			(Q3) & fulfillment of legal obligations, coordination (centralization) facilitates handling, good error culture benefits the company's image, more security (2x), high quality reports                    \\ \midrule
			(Q4)      & no disadvantages, efforts and discussion with 3rd parties                                   \\ \midrule
			(Q5)      & integration of and communication with the customer, best-case using CSAF, but it's a long way to go, 3rd parties desiring to be excluded, identifying all relevant parties \\ \bottomrule
		\end{tabular}%
	}
\end{table}

\begin{table}[!h]
\centering
\caption{Anonymized and aggregated answers we received during our \primarysurvey{}.}
\label{tab:appendix:isanswers}
\resizebox{\linewidth}{!}{%
\begin{tabular}{@{}p{0.13\linewidth}p{0.85\linewidth}@{}}
\toprule
\textbf{Question}    & \textbf{Answers}                                                                                                                                             \\ \midrule
(Q1)      & Yes (2x), No (3x)                                                                                                                                   \\ \midrule
(Q2)      & Yes (6x), No (1x)                                                                                  \\ 
(Q2a)     & 3 years, \textgreater{}11 years, since 01.01.2023, 2024, hygiene, channel for external vulnerability reports, increase of security level, prevention (1x), professional handling of vulnerabilities, Unanswered (3x) \\
(Q2b)     & Unanswered (5x)                                                                                                                                     \\
(Q2c)     & Unanswered (5x)                                                                                                                                     \\ \midrule
(Q3) &
  lack of resources (2x), quality \& quantity of findings, costs, uncertain responsibilities within departments, low security awareness in the past, hackers' attention due to CVD launch, no particular issues (2x), time consumption of setting up the processes, Unanswered (1x) \\ \midrule
(Q4)      & Yes (3x), No (1x), Unanswered (1x)                                                                                                                  \\
(Q4a)     & Security Operations Center, Unanswered (4x)                                                                                                         \\ \midrule
(Q5)      & 11-100 (2x), 101-1000 (1x), Unanswered (2x)                                                                                                         \\
(Q5a)     & 20-40\% (1x), 60-80\% (1x), 80-100\% (1x), Unanswered (2x)                                                                                          \\
(Q5b)     & mixed (1x), good (2x), very good (1x), Unanswered (1x)                                                                                              \\ \midrule
(Q6)      & 10k - 100k EUR (1x), 1m EUR (1x), Unanswered (3x)                                                                                                         \\ \midrule
(Q7)      & analysis \& CVSSv3, based on business impact, based on severity of the vulnerability, individual decisions for all reports, Unanswered (1x)         \\ \midrule
(Q8)      & 10-30 days (2x), 60+ days (1x), Unanswered (2x)                                                                                                     \\ \midrule
(Q9) &
  more trust (3x) from partners(1x) \& clients, more security (6x), structured input channel (clarity), more legal certainty, competence gain within the company, more reputation, better error culture, high quality vulnerabilities, Unanswered (1x) \\ \midrule
(Q10)     & none, human resources, high efforts \& costs, hard to find specialized personnel, discussions with external service providers, Unanswered (3x)                                                   \\ \midrule
(Q11)     & none, identification of suitable contact persons or parties (2x), high workload, lack of responsibility, budget, demands from external service providers, Unanswered (1x)                                    \\ \midrule
(Q12) &
  quality of findings, ratio valid / invalid, number of submissions \& criticality, number of uncoordinated disclosures should be 0, number of fixed vulnerabilities, Unanswered (1x) \\ \midrule
(Q13) &
  diverse application security vulnerabilities, a bit of everything, open redirections, out-of-bounds, improper input validation, improper restriction of operations within memory buffer, exposure of sensitive information, integer overflow, heap-based buffer overflow, improper authentication, uncontrolled resource consumption, Unanswered (2x) \\  \bottomrule
\end{tabular}%
}
\end{table}

\end{document}